\documentclass[aps,twocolumn,superscriptaddress,preprintnumbers]{revtex4}
\usepackage{epsfig}
\usepackage{graphicx,amssymb,epsfig,amsmath}
\begin{document}

\title{Spin-torque driven ferromagnetic resonance in a nonlinear regime}
\author{W. Chen}
\affiliation{Department of Physics, New York University, New York,
NY 10003}
\author{G. de Loubens}
\affiliation{Department of Physics, New York University, New York,
NY 10003}
\author{J-M. L. Beaujour}
\affiliation{Department of Physics, New York University, New York,
NY 10003}
\author{J. Z. Sun}
\affiliation{IBM T. J. Watson Research Center, Yorktown Heights, NY
10598}
\author{A. D. Kent}
\affiliation{Department of Physics, New York University, New York,
NY 10003}
\date{September 14, 2009}

\begin{abstract}
Spin-valve based nanojunctions incorporating Co$|$Ni multilayers
with perpendicular anisotropy were used to study spin-torque driven
ferromagnetic resonance (ST-FMR) in a nonlinear regime.
Perpendicular field swept resonance lines were measured under a large
amplitude microwave current excitation, which produces a large angle
precession of the Co$|$Ni layer magnetization. With increasing rf
power the resonance lines broaden and become asymmetric, with their
peak shifting to lower applied field. A nonhysteretic step jump in ST-FMR voltage
signal was also observed at high powers. The results are analyzed
in in terms of the foldover effect of a forced nonlinear
oscillator and compared to macrospin simulations. The ST-FMR nonhysteretic
step response may have applications in frequency and amplitude 
tunable nanoscale field sensors.
\end{abstract}
\maketitle

Spin-torque driven ferromagnetic resonance (ST-FMR) has proven to be
a powerful technique to investigate magnetic anisotropy, damping
and spin-torque interactions in nanometer scale metallic ferromagnets
\cite{Tulapurkar2005,Sankey2006,Kupferschmidt2006,Wang2006}. The
method involves injecting an rf current into a laterally confined
giant magnetic resistance or magnetic tunnel junction device that
drives one of the magnetic layers into resonance. Homodyne mixing of
the rf current and device resistance oscillations results in a dc
voltage across the device.

In our earlier studies, ST-FMR was conducted on spin valve junctions
that incorporate Co$|$Ni multilayers that have perpendicular
magnetic anisotropy and a thick Co reference magnetic layer
\cite{Chen2008APL, Chen2008MMM}. The Co$|$Ni layer magnetic
anisotropy was measured with  rf currents in the sub-mA range, well
below the devices' spin-transfer switching threshold of $\simeq$ 8 mA.
In these studies the rf current was modulated on and off at a low frequency ($\sim$
kHz), and a lock-in amplifier was used to measure the mixing
voltage, $V$. Measurements were conducted in a linear response
regime in which the homodyne signal was proportional to the rf
power, i.e. where the field swept resonance line,
$V(H)/I_\mathrm{rf}^2$ was independent of $I_\mathrm{rf}$. Here
$I_\mathrm{rf}$ is the amplitude of the rf current. The resonance
lines were symmetric and magnetization precession angles were small
(typically around 2$^\circ$ and no more than 6$^\circ$).

\begin{figure}
\includegraphics[width=0.48\textwidth]{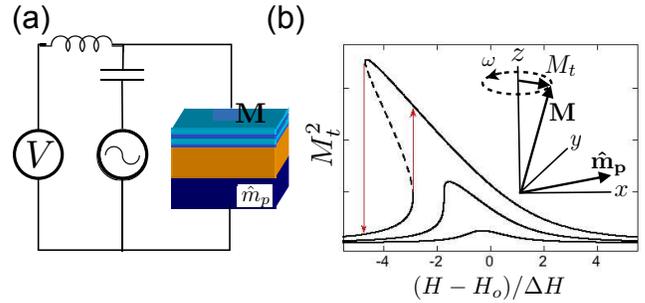}
\vspace{-1 cm}
\caption{(a) Sample layer structure and ST-FMR circuit. (b)
The foldover effect: precession amplitude vs field at fixed rf
frequency for three different normalized rf power levels,
$\beta=0.2\beta_c, \beta_c$ and $3\beta_c$  ($\beta_c=8/(3 \sqrt{3})$). In the
last case the curve is multivalued, to the left of the
resonance there is a low and high amplitude branch, leading to field sweep direction dependent 
hysteresis (red arrows). The dashed part
of the curve is an unstable oscillation. Inset shows the magnetization
configuration. The applied field is in the $z$ direction and the
magnetization of the Ni$|$Co layer precesses about this direction at
an angular frequency $\omega$ and with a transverse magnetization $M_t$.}
\label{Layout}
\end{figure}

In this Letter we present ST-FMR studies in a nonlinear regime,
where the amplitude of the rf current is much larger. The resonance
field and the linewidth of the Co$|$Ni layer have been determined
with increasing rf current amplitude and the rf current is not
modulated, enabling an investigation of the effect of field sweep
direction on the resonance lineshape.  We analyze these results with a
foldover theory of nonlinear oscillators that predicts a shift in
resonance field and a hysteretic response, with a lineshape that
depends on the field sweep direction (Fig. \ref{Layout}(b)). We also
compare our results to macrospin simulations.

Pillar junctions with a lateral dimension of 50$\times$150 nm$^2$
(Fig. \ref{Layout}(a)) were patterned on a silicon wafer using a
nanostencil process \cite{Sun2002}. Junctions were deposited using
metal evaporation with the layer structure $\parallel1.5$ nm Cr$\mid
100$ nm Cu$\mid 20$ nm Pt$\mid 10$ nm Cu$\mid$ [0.4 nm Co$\mid$ 0.8
nm Ni]$\times$ 3 $\mid 10$ nm Cu$\mid 12$ nm Co$\mid 200$ nm
Cu$\parallel$, where the thin (free) layer is composed of a Co$|$Ni
multilayer and the thick (fixed) layer is pure Co.

ST-FMR measurements were conducted with a magnetic field applied
nearly perpendicular (within $\sim2^\circ$) to the sample surface at room
temperature. An rf current generated by a high frequency source is
amplified by 20 dB and is added to a dc current using a bias-T (Fig.
\ref{Layout}(a)). Positive dc currents are defined such that
electrons flow from the free layer to the fixed layer. 
The magnetoresistance is 2.1$\%$ and the pillar
resistance is 1.53 $\Omega$ in the parallel state (Fig. 1(b) of Ref.
\cite{Chen2008APL}). The inset of Fig. \ref{Layout}(b) shows the magnetization configuration.
With an applied field greater than 0.3 Tesla, the Co$|$Ni
multilayer is magnetized along $z$, perpendicular to the plane of the
layers. However, since the Co layer has a large
magnetization, $\mu_0 M_s \simeq 1.5$ T, it remains largely magnetized in
plane, as represented by the vector $\bf{\hat{m}}_P$. rf currents
cause the magnetization of the Co$|$Ni multilayer, $\bf{M}$, to
precess with a transverse amplitude $M_t$ at an angular frequency $\omega$.

\begin{figure}[t]
\includegraphics[width=0.48\textwidth]{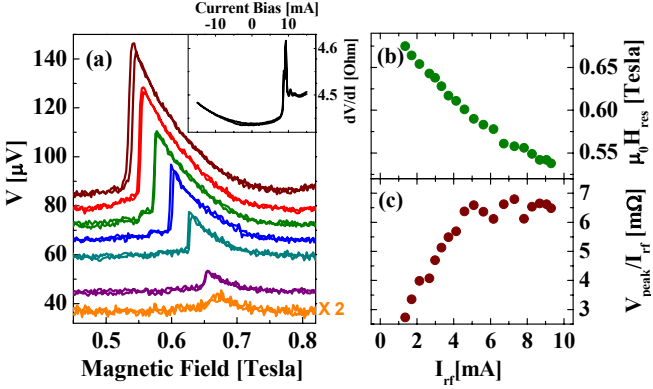}
\vspace{-1 cm}
\caption{(a) ST-FMR signal as a function of the applied field
at different rf currents levels for $f=16$ GHz.
$I_\text{rf}$ was 1.4, 2.1, 3.3, 4.6, 6.2, 7.8, and 9.0 mA
respectively. Each adjacent curve is offset by 10 $\mu$V. The
resonance at $I_\text{rf}$=1.4 mA is amplified by a factor of 2
above the background of 37 $\mu$V. Inset: dV/dI \textit{vs} I of
at $\mu_0H$=0.675 T, which is the
resonance field at  $f$=16 GHz in the linear regime. (b) Magnetic field where the step jump
occurs \textit{vs} $I_\text{rf}$. (c) $V_\text{peak}/I_\text{rf}$ \textit{vs} $I_\text{rf}$.
The peak voltage is measured with respect to the baseline background
voltage.} \label{Fig:Resonance}
\end{figure}

ST-FMR was conducted with a constant rf power, and a dc voltmeter
was used to measure the mixing voltage. The applied field was swept
up and down at a rate 0.12 T/min. Fig. \ref{Fig:Resonance}(a) shows an example of such measurements.
In this case, the rf frequency $f$ is 16 GHz and $I_\text{rf}$ ranges between 1.4 and
9.0 mA. Each curve is offset by 10 $\mu$V for clarity. At
the lowest rf amplitude ($I_\text{rf}$=1.4 mA), there is a
37 $\mu$V non-resonant background, which we associate with nonlinearities in
our measurement circuit. This offset varies with the rf amplitude, which we attribute to other parasitic rectification processes in our measurement circuits, such as slightly non-ohmic contacts and junction heating effects.
At $I_\mathrm{rf}$=1.4 mA we are in the linear response regime, the lineshape is symmetric, and the resonance field is 0.675 T. The critical current for spin-torque-induced
dynamics at this field is $\sim$8.5 mA (inset of
Fig. \ref{Fig:Resonance}(a)). The lineshape begins to become asymmetric
at $I_\mathrm{rf}$=2.1 mA. Starting from $I_\mathrm{rf}$=3.3 mA a
step jump of the mixing voltage occurs on the lower-field side of
the resonance. The maximum ST-FMR voltage increases as
$I_\mathrm{rf}$ increases, and the step-jump field continues to decrease.

The shift of the resonance peak with increasing power can be
understood in terms of forced oscillator in which the frequency of
oscillator depends on the oscillation amplitude, i.e. the system is
a driven anharmonic oscillator. The micromagnetic energy of the free
layer can be written as:
\begin{equation}
E= -\mu_0 M_0 m_z H + {1 \over 2} \mu_0 M_0 M_{\mathrm{eff}} m_z^2,
\label{Eq:Energy}
\end{equation}
where $m_z=M_z/M_0$ and $M_{\mathrm{eff}}= M_{0} - 2K_1/(\mu_0 M_{0})$
is the effective easy-plane anisotropy. The first term is the Zeeman
energy and the second term is written in terms of
$M_{\mathrm{eff}}$, which contains contributions due to shape anisotropy
($M_0$) and the layer perpendicular magnetic
anisotropy ($K_1$). The resonance condition in this geometry 
(applied field along $z$) is given by \cite{Kittel}, 
$2\pi f=\gamma \mu_0(H_{\mathrm{res}}-M_{\mathrm{eff}}m_z)$,
where $\gamma$ is the gyromagnetic ratio. For a fixed frequency the
resonance field is given by:
\begin{equation}
H_{\mathrm{res}}=\frac{2\pi f}{\mu_0 \gamma}+M_{\mathrm{eff}}m_z
\label{Eq:Hres}
\end{equation}
As the precession amplitude $m_t=M_t/M_0$ increases, the longitudinal component of
magnetization, $m_z \simeq1-m_t^2/2$, decreases resulting in a shift of the
resonance to lower fields ($\mu_0M_{\mathrm{eff}}\simeq$ 0.2 T for our
Co$|$Ni multilayer), as observed in our experiments.

For small rf current amplitudes the lineshape is Lorentzian:
\begin{equation}
m_t^2= \frac{a I_\mathrm{rf}^2}{(H-H_0)^2+(\Delta H)^2}
\end{equation}
with a half width at half maximum of $\Delta H=\alpha
\omega/(\gamma \mu_0)$, where $a$ is a constant that depends on the
spin-torque per unit current and the magnetization direction of the
fixed magnetic layer, $\hat{m}_P$. $\alpha$ is the Gilbert damping constant.
With increasing precession amplitude one finds \cite{Anderson1955,Schlomann1959, Landau1976, Fetisov1999} to first order in $m_t^2$:
\begin{equation}
\tilde{m}_t^2= \frac{1}{(\tilde{\epsilon}+\beta \tilde{m}_t^2)^2+1}.
\label{Eq:FoldoverForm}
\end{equation}
Here $\tilde{m}_t=m_t/m_{tmax}$ is the transverse component of
magnetization normalized to the maximum precession amplitude,
$m_{tmax}^2= aI_\mathrm{rf}^2/(\Delta H)^2$;
$\tilde{\epsilon}=(H-H_0)/\Delta H$.  The resonance peak position is given by:
$\tilde{\epsilon}=-\beta=-a M_{\mathrm{eff}}
I_{\mathrm{rf}}^2/[2(\Delta H)^3]$. So the resonance peak
position is predicted to shift with the rf power,
$I_\mathrm{rf}^2$. 

\begin{figure}
\includegraphics[width=0.48\textwidth]{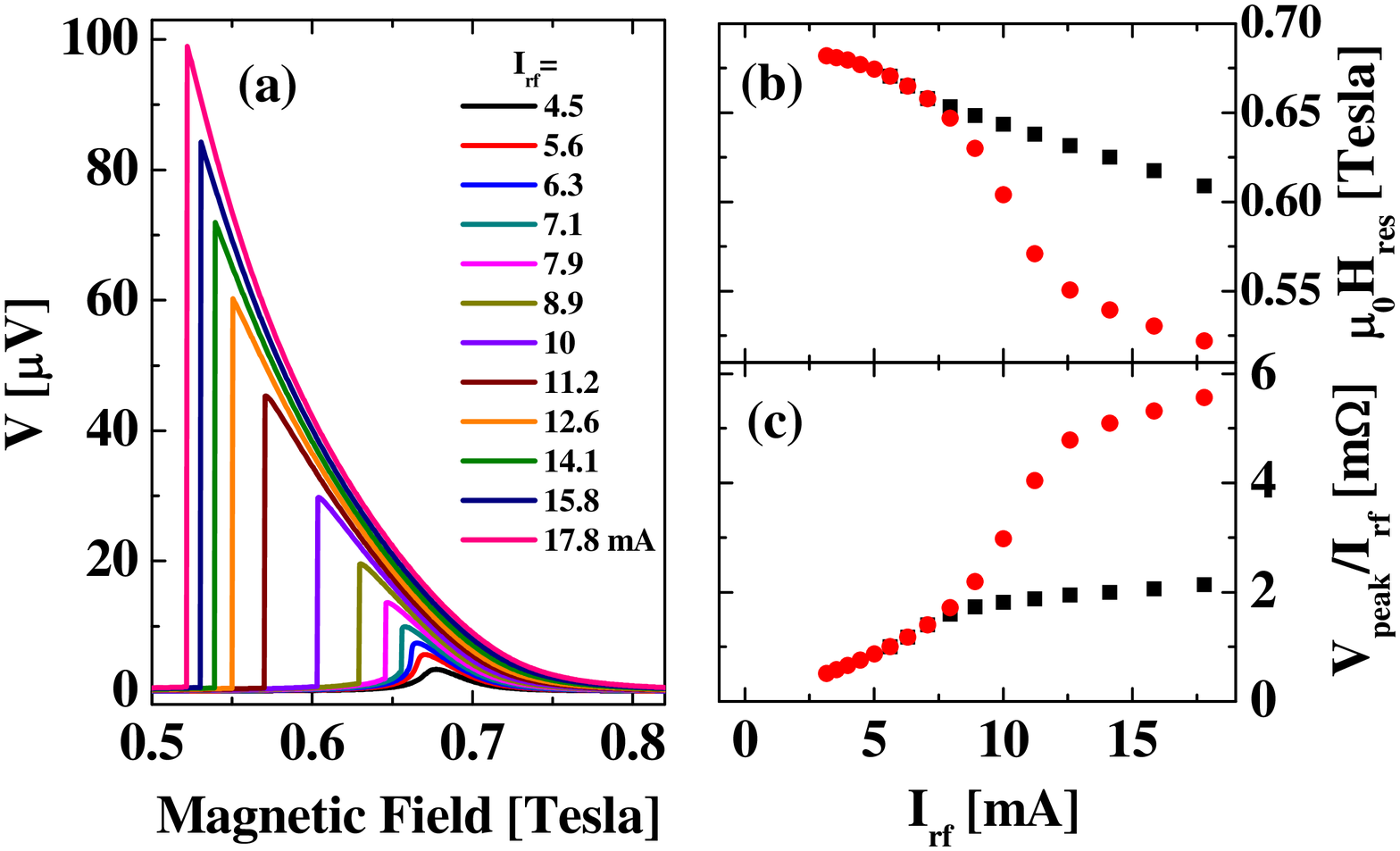}
\vspace{-1 cm}
\caption{Macrospin micromagnetic simulations of ST-FMR.
(a) Voltage versus applied field for various rf currents,
showing the high amplitude branch of the curve. (b) Resonance
peak position and (c) $V_\mathrm{peak}/I_\mathrm{rf}$ vs $I_\mathrm{rf}$. The black
squares refer to the low amplitude branch and the red circles refer to the high amplitude branch
of the resonance curve.}
\label{Fig:Micromag}
\end{figure}

Fig. \ref{Layout}(b) shows a plot of $\tilde{m}$ versus field for
three different rf amplitudes. There is a critical value of
$\beta_c=8/(3\sqrt{3})\simeq1.54$, which corresponds to the foldover threshold
for the excitation amplitude $I_\mathrm{rf}$, for which the left side of the resonance curve
has a vertical tangent. For sufficiently large current,
$\beta>\beta_c$, the resonance curve is multivalued, with two
stable/metastable branches differing in amplitude that overlap on
the low field side of the resonance in Fig. \ref{Layout}(b). The
dashed part of the curve in Fig. \ref{Layout}(b) is an unstable oscillation. As a result
hysteresis is predicted in field swept measurements, as indicated by
the red arrows in the figure. Hysteresis begins when the shift of
the resonance peak is greater than approximately the resonance
linewidth ($H_\mathrm{res}-H_0$=$\beta_{c}\Delta H$).

The experimental data (Fig.~\ref{Fig:Resonance}(a)) have features
qualitatively consistent with the foldover model described above:
(1) The resonance lines become asymmetric with increasing power; (2) There is a step jump in ST-FMR
voltage above a critical rf amplitude; and (3) The resonance peak
position decreases with increasing rf power (Fig.~\ref{Fig:Resonance}(b)).
However, there are discrepancies with
the model, particularly the fact that little hysteresis is observed
at large rf amplitudes, when the resonance shift far exceeds the resonance
linewidth, $\mu_0\Delta H=0.02$ T. Also, the resonance field shifts
nearly linearly with increasing rf current. 

To further understand the data we have conducted macrospin
simulations of the magnetization dynamics with an applied rf current.  We
take the micromagnetic energy given by Eq.~\ref{Eq:Energy} and an angular
dependence of the spin-torque and magnetoresistance determined using continuous random
matrix theory (CRMT) for our device layer stack \cite{Rychkov2009}. This model has been found
to give spin-torques in agreement with those found in our experiments conducted in the linear
response regime \cite{Chen2008APL}. The torque is parameterized by 
$\tau_\mathrm{ST}=I k(\theta) (\hbar/ 2e) \hat{m} \times (\hat{m} \times \hat{m}_p)$
with $k(\theta)=(a+b \sin\theta)/(c + d \cos\theta)$ ($a=1.98,\;b=7.11,\;c=28.74,\;
d=25.95$). The results of our simulations are shown in Fig.~\ref{Fig:Micromag}. The high amplitude branch, with extends to lower field,
is shown in Fig.~\ref{Fig:Micromag}(a). The resonance lines shapes are similar to those observed in experiment. Fig.~\ref{Fig:Micromag}(b) shows the shift of the peak position for both of high and low amplitude branches as well as the peak voltage. For large $I_\mathrm{rf}$ the resonance peak position shifts linearly with  $I_\mathrm{rf}$ and $V_\mathrm{peak}/ I_\mathrm{rf}$ saturates at large $I_\mathrm{rf}$  (Fig.~\ref{Fig:Micromag}(c)), as seen in experiment. The main discrepancy is that larger rf currents are needed to produce the resonance shifts observed.

It also appears that the resonance at high rf currents follows the high amplitude branch independently of the field sweep direction. The evidence for this is the following. First, little to no hysteresis is observed. Second, the resonance shift far exceeds the linewidth as well as the resonance shift found for the low amplitude branch in micromagnetic simulations (Fig. \ref{Fig:Micromag}(b)). We suspect that there are thermally driven transitions between these dynamical states that excite the high amplitude branch and reduce the hysteresis. The rf current may also heat the device reducing the magnetization of the free layer, increasing thermal fluctuations and leading to a larger shift of the resonance peak position with rf current than seen in the macrospin model. We also note that our analysis and macrospin simulations assume that the damping does not depend on precession amplitude. However, it has been noted that the damping may be nonlinear and this may play a role in understanding the large amplitude ST-FMR driven magnetization dynamics \cite{Tiberkevich2007}.

It is interesting to estimate the maximum precession angle, $\theta_\mathrm{max}$. This can be done using the voltage peak as well as maximum resonance shift. The peak mixing voltage is given by $V$=$\frac{1}{4}\Delta R I_\text{rf}\sin\eta\sin\theta_\mathrm{max}$, where $\Delta R$ is the junction magnetoresistance and $\eta$ is the angle between the layers with no rf current applied ($\eta \simeq 70^\circ$, in these experiments). The precession angle can also be estimated from the shift of the resonance field using Eq. \ref{Eq:Hres}. Both approaches give 
$\theta_\mathrm{max}\sim$65$^\circ$at $I_\text{rf}$=9.3 mA, the largest rf current applied in the junction.

In sum, these results illustrate that large rf currents can drive nonlinear
magnetization dynamics, characteristic of any driven anharmonic oscillator. The observed nonhysteretic
step response may prove useful for rf frequency and amplitude tunable nanometer scale field sensors. It will also be interesting to compare these results to full micromagnetic simulations as well as to further explore the role of thermal fluctuations on ST-FMR foldover phenomena.

We thank A. N. Slavin for useful discussions and X. Waintal for the CRMT code used in our macrospin simulations. This research is supported by NSF-DMR-0706322 and the ARO-W911NF0710643.

\end{document}